\documentstyle [prl,aps] {revtex}

\begin{document}

\draft

\title{On the Numerical Solution of the Time Dependent Schr\"odinger
Equation }

\author{R. Sch\"afer}
\address{
University of Siegen,
         Fachbereich Physik,  D-57068 Siegen, Germany,
}
\author{R. Blendowske}
\address{ Leica Mikroskopie und Systeme GmbH,
          D-35530 Wetzlar, Germany}

\date{September 1994}

\maketitle

\begin{abstract}
An  algorithm for the numerical solution of the
Schr\"odinger equation in the case of a time dependent potential is proposed.
Our simple modification upgrades the well known  method  of Koonin
while negligibly increasing the
computing time. In the presented test the
accuracy is enhanced by up to an order of magnitude.
\end{abstract}

\pacs{02.60.+y, 25.85.-w}

The microscopic description of many--body systems like atoms or nuclei
is based on a many--body Hamiltonian. The related wave functions
are given by Slater determinants for fermions. In the case
of time dependent processes like  atomic or nuclear collisions, nuclear
fission or fusion, however, the situation  is too complex due to the great
number of degrees of freedom. Therefore, in most cases, a collective
coordinate is introduced according to the essential physical
properties of the considered system. This procedure leads to   a
macroscopic model with one degree of freedom,  which is governed by an
effective one--body  Schr\"odinger equation including a time dependent
potential in the considered examples \cite{1}.
The well known coordinate representation reads
\begin{equation}
i\hbar\; \frac{\partial}{\partial t}\; \phi \left(  {\bf r},t \right)
=
\left[ -\frac{\hbar^2}{2 \: M}\; \nabla^2 \;
+ V \left(  {\bf r},t\right)  \right]  \; \phi \left(  {\bf r},t \right)
\; .
\label{eq:a}
\end{equation}
 A momentum dependent potential $V$,
leading to an integro-differential equation, is excluded here. In
almost all practical cases eq. (\ref{eq:a}) has to be evaluated
numerically. The standard  technique essentially consists of
the following two steps: first apply a suitable scheme for the space
discretization and then perform the time integration.

The algorithm presented here is only related to the second step:
The time integration in the case of an explicitly time dependent potential
$V({\bf r},t)$. Our  proposed convenient modification upgrades the
standard method to a much more efficient version while negligibly
increasing the computing time.

Our experience is due to the description of ternary fission, i.e. a
fission process accompanied by the emission of an $\alpha$--particle,  with
results to be published elsewhere. However, the presented algorithm
might be of broader interest, not necessarily restricted to  this
nuclear physics theme.

In this letter  we will discuss the   application for the case of a
cylindrically
symmetric, time dependent potential, essentially following Koonin's
procedure \cite{2}, and present a test using  an analytically
solvable example.

As usual, in a cylindrically symmetric  case,
the angle dependence on $\varphi$ of
the wave function is separated as
\begin{equation}
\phi = \frac{1}{\sqrt{2 \pi }} \; \psi(\rho,z,t) \;\exp(i\mu\varphi)
\; ,
\end{equation}
and  a grid is defined by
\begin{eqnarray}
\rho_j  &=& (j-\frac{1}{2})\; \Delta \rho \:,\quad j=1, \ldots,
N_{\rho} \; ,
\nonumber \\
z_k     &=&  k \; \Delta z \:,\quad k=-N_z, \ldots, N_z
\; .
\label{eq:4}
\end{eqnarray}
For simplification we introduce now
\begin{equation}
g_{j,k} = \sqrt{\rho_j} \; \psi_{j,k}
\end{equation}
leading to the space--discretized equation
\begin{equation}
i \hbar \; \frac{\partial}{\partial t} \; g_{j,k}
=
(H g)_{j,k} = (v g)_{j,k} + (h g)_{j,k}
\; ,
\label{eq:10}
\end{equation}
 where the Hamiltonian $H$ is split in a
``vertical'' ($v$)  and a ``horizontal'' ($h$) part
\begin{eqnarray}
(v g)_{j,k} & = &
 - \frac{\hbar^2}{2 M}\;
\frac{\;c_j\;g_{j+1,k}\; -2 \;g_{j,k}\; + c_{j-1}\;g_{j-1,k}\;}{\Delta
\rho\; ^2}
 +\; U_{j,k} \; g_{j,k}
\; ,
\nonumber\\
(h g)_{j,k} & = &
- \frac{\hbar^2}{2 M}\;
\frac{\; g_{j,k+1} \; - 2\;g_{j,k}\; + \;g_{j,k-1}\;}{\Delta z\; ^2}
+\; U_{j,k} \; g_{j,k}
\; ,
\end{eqnarray}
with the abbreviations
\begin{eqnarray}
   c_j & = & \frac{j}{\sqrt{j^2- 1 / 4}}
\; , \nonumber\\
U_{j,k}& = & \frac{1}{2} V_{j,k} + \frac{\mu^2 \hbar^2}{4 M \rho_j^2}
\; .
\end{eqnarray}
Up to this point no changes are made in comparison to Koonin's
algorithm. We  solve eq. (\ref{eq:10})  iteratively, with  times
 $t_{n} = n  \Delta t$, using a Taylor expansion of $g_{jk}(t_n)=
g^{(n)}_{jk}$,  up
to and including $(\Delta t)^2$ and  obtain
\begin{eqnarray}
\lefteqn{
g^{(n+1)}  =  \left( \;1-\frac{i}{\hbar}\;H \;(\Delta t)^1 \right.
}
\nonumber\\
&& \left.
- \: \frac{1}{2\hbar^2}\; H ^2 \; (\Delta t)^2
\;
-\frac{i}{2\hbar}\;\dot{V}  \; (\Delta t)^2
+{\cal O} \left(   \Delta t \;^3 \right) \right) \quad g^{(n)}
\; ,
\label{eq:20}
\end{eqnarray}
where all terms on the right hand side are evaluated at time $t_n$.
The time derivative
$ \dot{V}_{j,k} = \partial V_{j,k} / \partial t$ is
included to ensure consistency up to second order contributions.

Inserting $g^{(n)}$ on the right hand side is not an advisable method
to determine $g^{(n+1)}$, because of serious numerical instabilities.
Therefore,  an  ``alternating direction implicit method'' ~\cite{3} is
used, leading to a modified version  of Koonin's algorithm
\begin{eqnarray}
g^{(n+1)}&=&
\left( \; 1 + \frac{i}{2\hbar} \;v \;\Delta t +
\frac{i}{8\hbar} \:\dot{V}\:(\Delta t)^2 \;\right)  ^{-1}  \nonumber\\
&& \left(  1 - \frac{i}{2\hbar} \; h\; \Delta t -
\frac{i}{8\hbar} \:\dot{V}\:(\Delta t)^2 \right)   \nonumber \\
&& \left( \; 1 + \frac{i}{2\hbar} \;h\; \Delta t +
\frac{i}{8\hbar} \: \dot{V}\:(\Delta t)^2 \;\right) ^{-1}\nonumber \\
&&\left( 1 - \frac{i}{2\hbar}\; v\;\Delta t  -
\frac{i}{8 \hbar} \:\dot{V}\:(\Delta t)^2 \right) \quad g^{(n)}
\; ,
\label{eq:21}
\end{eqnarray}
where again all terms on the right hand side are evaluated at time $t_n$.
Our result  may be verified by an expansion in terms of  $\Delta t$ up to
second order. There are no restrictions on the
commutation of  $[v,h]$. By neglecting terms dependent on $\dot{V}_{j,k}$,
our  expression reduces to the one used by Koonin et al. in~\cite{2}.

For numerical convenience we now define  $w^{(n)}$ by
\begin{equation}
\left( \; 1 + \frac{i}{2\hbar} \; h\;\Delta t  +
\frac{i}{8\hbar} \:\dot{V}\:\ (\Delta t)^2 \;\right)  \;w^{(n)}
=
\left( \; 1 - \frac{i}{2\hbar} \; v \;\Delta t  -
\frac{i}{8\hbar} \:\dot{V}\: (\Delta t)^2 \;\right) \; g^{(n)}
\label{eq:24}
\end{equation}
leading to
\begin{equation}
\left( \; 1 + \frac{i}{2\hbar}\; v \;\Delta t  +
\frac{i}{8\hbar} \:\dot{V}\:(\Delta t)^2 \;\right)
\;g^{(n+1)}
=
\left( \; 1 - \frac{i}{2\hbar}\; h \; \Delta t  -
\frac{i}{8 \hbar} \:\dot{V}\:(\Delta t)^2 \;\right) \; w^{(n)}
\; .
\label{eq:25}
\end{equation}

In order to determine $w^{(n)}$, and  thereafter $g^{(n+1)}$,  one has to
invert tridiagonal matrices only, which  can be performed by a suitable
algorithm \cite{4}.

A central point for the efficiency of our modification is that the
determination of
\begin{equation}
\dot{V}^{(n)}_{j,k}   =
\frac{V^{(n)}_{j,k} - V^{(n-1)}_{j,k} }{\Delta t}
\label{eq:22}
\end{equation}
up to order $(\Delta t)^0$
poses no appreciable
effort. Anyway, the values of $V_{j,k}$ have to be calculated  for
all time steps in order to determine $v$ and $h$.
It is deemed unsuitable to compensate this by determining $v$ and
$h$  at an intermediate  time step  $t + \Delta t /2$
 because one has to calculate the potential at this additional
time step, as well.
In many cases this produces appreciable effort. For example, in our
calculations concerning the ternary fission  it took approximately the
same time to calculate $V^{(n)}$ for one value of $n$
on the whole grid  as  to
perform one iteration step from $g^{(n)}$ to $g^{(n+1)}$.

To test our algorithm, with regard to the modification due to keeping
terms dependent on $\dot{V}$, we consider
the following, explicitly time dependent potential
\begin{equation}
V \left(  {\bf r},t \right)  = \frac{1}{2}\: M \omega ^2 \, {\bf r}\;^2
-2 \hbar \omega^2 \, t
\; .
\label{eq:26}
\end{equation}
At time $t=0$ this is the potential of the harmonic oszillator with
frequency $\omega$. For the initial wave function we choose an
eigenfunction of this harmonic oszillator
\begin{equation}
\phi \left(  {\bf r},t=0 \right) =  c_\ell \;H_\ell (\sqrt{\beta}\; z)  \;
\sqrt{\beta /\pi} \;
\exp \left( - \: \frac{\beta}{2} \left[\rho^2 + z^2 \right] \right)
\; ,
\label{eq:27}
\end{equation}
where
$ \beta =  M \omega / \hbar$,
$\ell \in \left\{0,1,2,\ldots \right\}$,
$H_\ell =$ Hermite polynomial,
and $c_\ell = \mbox{constant of normalization}$.

The analytical solution is given by
\begin{equation}
\phi \left(  {\bf r},t \right)   =
\phi \left(  {\bf r},t=0 \right)   \:
\exp \left(
\displaystyle{ - \, i \omega \left[\ell+ 3/2\right] t+i \omega ^2 \,
t^2 }
\right)
\; .
\label{eq:28}
\end{equation}
The results are displayed in Table \ref{table:a} for different radial
quantum \mbox{numbers $\ell$}
(different numbers of nodes of the wave function).
Both parts of the wave function---real and imaginary---are
considered separately. We use $\omega = 2$, $\Delta t = 1/60$ with 60
time steps and grid parameters $N_{\rho}=N_{z}=64$ and $\Delta \rho =
\Delta z = 0.25$.   The enhancement in accuracy becomes as large as
an order of magnitude.
The norm of the wave function is (nearly) conserved, since the
algorithm---modified or not---is nearly unitary \cite{2}.

We note that the considerations for a one dimensional time dependent
potential are quite the same. One obtains
\begin{eqnarray}
\lefteqn{
\phi \left(  x , t + \Delta t \right)
 =
\left( \; 1 + \frac{i}{2\hbar} \: H  \:  \Delta t +
\frac{i}{4\hbar} \:\dot{V}
 \:  (\Delta t)^2 \;\right)  ^{-1}
}
\nonumber \\
&&
\left( \; 1 - \frac{i}{2\hbar} \: H  \: \Delta t
-  \frac{i}{4\hbar} \:\dot{V}  \: (\Delta t)^2 \;\right)
\phi \left(  x,t \right)   +  {\cal{O}} \;\left(  \;\Delta t\;^3 \;\right)
\; .
\nonumber \\
&&\label{eq:32}
\end{eqnarray}
Again the calculation  of $\dot{V} \left(  x,t \right) $
poses nearly no effort. In contrast to the
ordinary method, where the $\dot{V}$ terms are neglected~\cite{4}, the
above expression is accurate up to order $ (\Delta t)^2$.
The approximation (\ref{eq:32}) is unitary; it automatically maintains
the normalization of $\phi$. We note that no
alternating direction method is needed in the one dimensional case.

We summarize: In the case of a time dependent potential the standard
algorithm of Koonin can easily be modified to obtain a greater accuracy.
Therefore, terms depending on $\dot{V}$ have to be included to get a
consistent time  expansion up to order $(\Delta t)^2$. The related algorithm
has been shown to be superior in the presented numerical test.

\newpage

\begin{table}
\caption{Results of the numerical test. The relative error of the real
and imaginary part of the wave function is displayed for  different
numbers of nodes. The modification of the Koonin's standard  algorithm
lead to an enhancement in the accuracy of up to an order of magnitude.}
\label{table:a}
\begin{tabular}{c c c c c }
nodes   &  \multicolumn{2}{c} {Relative error of} &
           \multicolumn{2}{c} {Relative error of}\\
        &  \multicolumn{2}{c} {the real part:}     &
           \multicolumn{2}{c} {the imaginary part:} \\
        & standard & modified & standard & modified \\
\hline
$\ell=0$ &  $ 10 \% $ & $ 0.4\% $ &
            $ 4 \%  $ & $ 0.2\% $ \\
$\ell=1$ &  $ 10 \% $ & $ 0.6\% $ &
            $ 4 \%  $ & $ 0.4\% $ \\
$\ell=2$ &  $ 1 \%  $ & $ 0.3\% $ &
            $ 40 \% $ & $ 5\%   $ \\
$\ell=3$ &  $ 20 \% $ & $ 4\%   $ &
            $ 1.5\% $ & $ 0.6\% $ \\
\end{tabular}
\label{table:1}
\end{table}


\begin{references}
\bibitem{1} K.~Langanke, J.~A.~Maruhn, S.~E.~Koonin,
            Computational Nuclear Physics 2, Nuclear Reactions
            (Springer-Verlag, New York 1993)
\bibitem{2} S.~E.~Koonin, K.~T.~R.~Davies, V.~Maruhn-Rezwani, H.~Feldmeier,
            S.~J.~Krieger, J.~W.~Negele,
            Phys. Rev. {\bf C15}  (1977) 1359
\bibitem{3} R.~S.~Varga,  Matrix Iterative Analysis (Prentice-Hall,
            Englewood Cliffs 1962)
\bibitem{4} W.~H.~Press, B.~P.~Flannery, S.~A.~Teukolsky, W.~T.~Vetterling,
            Numerical Recipes  (Cambridge University Press 1989)
\end{references}
\end{document}